\documentstyle[amssymb,aps]{revtex}
\begin{document}
\title{Faithful remote state preparation using finite classical bits and a
nonmaximally entangled state}
\author{Ming-Yong Ye, Yong-Sheng Zhang\thanks{%
Electronic address: yshzhang@ustc.edu.cn}, Guang-Can Guo\thanks{%
Electronic address: gcguo@ustc.edu.cn}}
\address{Key Laboratory of Quantum Information, University of Science and Technology\\
of China, CAS, Hefei 230026, People's Republic of China\bigskip \bigskip }
\maketitle

\begin{abstract}
\baselineskip12pt We present many ensembles of states that can be remotely
prepared by using minimum classical bits from Alice to Bob and their
previously shared entangled state and prove that we have found all the
ensembles in two-dimensional case. Furthermore we show that any pure quantum
state can be remotely and faithfully prepared by using finite classical bits
from Alice to Bob and their previously shared nonmaximally entangled state
though no faithful quantum teleportation protocols can be achieved by using
a nonmaximally entangled state.

PACS number(s): 03.67.HK, 03.65.Ud
\end{abstract}

\baselineskip12pt

\section{Introduction}

The question ``What tasks may be accomplished using a given physical
resource?'' is of fundamental importance in many areas of physics \cite
{Nielsen/PRL83/436,quant/0205057}. Remote state preparation (RSP) \cite{Lo
/PRA62/012313} and quantum teleportation \cite{Bennett/PRL70/1895} answer
partly this question. Both protocols use classical communication and the
previously shared entangled state to prepare a quantum state in a remote
place. The differences between them are as follows. First, in RSP the sender
(Alice) knows the state she wants Bob to prepare while in quantum
teleportation Alice need not know the state she wants to send. Second, in
RSP, the required resource can be traded off between classical communication
cost and entanglement cost while in quantum teleportation, two bits of
forward classical communication and one ebit of entanglement (a maximally
entangled pair of qubits) per teleported qubit are both necessary and
sufficient, and neither resource can be traded off against the other \cite
{Bennett/PRL87/077902}. H. K. Lo has shown that for some special ensembles
of states, RSP requires less asymptotic classical communication than quantum
teleportation \cite{Lo /PRA62/012313}. Bennett {\it et al.} have shown that
in the high-entanglement limit the asymptotic classical communication cost
of remotely preparing a general qubit is one bit, which is also necessary by
causality \cite{Bennett/PRL87/077902}. Recently, Berry {\it et al.} have
shown it is possible to remotely prepare an ensemble of noncommutative mixed
states by using communication that is equal to the Holevo information for
this ensemble \cite{Berry/PRL90/057901}. Bennett {\it et al. }\cite
{Bennett/PRL87/077902} and Devetak {\it et al. }\cite{Devetak/PRL87/197901}
have also investigated low-entanglement remote state preparation which uses
more classical bits but less entanglement bits. The results were achieved
asymptotically.

Different from the above mentioned researchers, some others investigated
faithful and nonasymptotic remote state preparation \cite
{Pati/PRA63/014302,Leung/PRL90/127905,peng,Zeng/PRA65/022316}. Pati has
shown that a qubit chosen from equatorial or polar great circles on a Bloch
sphere can be remotely prepared with one classical bit from Alice to Bob if
they share one ebit of entanglement \cite{Pati/PRA63/014302}. Leung and Shor
have proved that if faithful RSP protocols without back communicating can
transmit generic ensembles and are oblivious to Bob, they can be modified to
become protocols oblivious to Alice. This indicates that they use at least
as much classical communication as that in quantum teleportation \cite
{Leung/PRL90/127905}.

In this paper we generalize Pati's RSP protocol \cite{Pati/PRA63/014302} to
nonmaximally entanglement and higher-dimensional case. In Sec. II, we
present a necessary and sufficient condition of a general RSP protocol,
similar to that proposed by Leung and Shor \cite{Leung/PRL90/127905} and
that by Hayashi {\it et al.} \cite{quant-ph/0205009}. Then, we investigate
RSP protocols using minimum classical bits. In Sec. III, we investigate RSP
protocols that do not use minimum classical bits, and prove that any pure
quantum state can be remotely prepared by using finite classical bits and
the previously shared nonmaximally entangled state. In Sec. IV, we shall
summarize and draw some conclusions.

\section{RSP achieved by using Minimum Classical Bits}

A general Pati's RSP protocol is characterized as follows. Alice and Bob
share an entangled state in two $d$-dimensional systems 
\begin{equation}
\left| \Psi _{AB}\right\rangle =\sum_{i=0}^{d-1}\alpha _i\left|
i\right\rangle \left| i\right\rangle ,\ \alpha _i>0,\sum_{i=0}^{d-1}\alpha
_i^2=1,  \eqnum{1}
\end{equation}
where $\left\{ \left| i\right\rangle \right\} _{i=0}^{d-1}$ forms an
orthonormal basis of $d$-dimensional Hilbert space. Alice wants Bob to
prepare a state $\left| \Phi \right\rangle $ which is known to her. She
performs a positive operator valued measurement (POVM) on her system $A$
with measurement operators that depend on the state $\left| \Phi
\right\rangle $. When Alice gets the result $m$ with the probability $%
p_m\left( \Phi \right) ,$ Bob's system $B$ will be in the state $\rho
_m\left( \Phi \right) =\left| \Phi _m\right\rangle \left\langle \Phi
_m\right| $. Alice sends the measurement result $m$ to Bob and Bob performs
the corresponding unitary operation $u_m,$ which change his system into the
state $\left| \Phi \right\rangle $ $\left( u_m\left| \Phi _m\right\rangle
=\left| \Phi \right\rangle \right) $. It is necessary that $u_m$ is
independent of $\left| \Phi \right\rangle $ and that $\rho _m\left( \Phi
\right) $ is a pure state. For Bob, before he receives the result $m,$ his
system is 
\begin{equation}
\sum_{m=0}^{n-1}p_m\left( \Phi \right) \left| \Phi _m\right\rangle
\left\langle \Phi _m\right| =\sum_{i=0}^{d-1}\alpha _i^2\left|
i\right\rangle \left\langle i\right| .  \eqnum{2}
\end{equation}
Substituting $u_m\left| \Phi _m\right\rangle =\left| \Phi \right\rangle $
into Eq. (2), we can obtain 
\begin{equation}
\sum_{m=0}^{n-1}p_m\left( \Phi \right) u_m^{\dagger }\left| \Phi
\right\rangle \left\langle \Phi \right| u_m=\sum_{i=0}^{d-1}\alpha
_i^2\left| i\right\rangle \left\langle i\right| .  \eqnum{3}
\end{equation}
Eq. (3) is a necessary condition for such RSP protocols. It is also a
sufficient condition \cite{Leung/PRL90/127905,quant-ph/0205009}, because
Alice only needs to apply a measurement on her system $A$ with POVM operators
\[
\{M_m=p_m\left( \Phi \right) \left( \sum_{i=0}^{d-1}\frac 1{\alpha _i}\left|
i\right\rangle \left\langle i\right| \right) \rho _m^T\left( \Phi \right)
\left( \sum_{i=0}^{d-1}\frac 1{\alpha _i}\left| i\right\rangle \left\langle
i\right| \right) \}_{m=0}^{n-1},
\]
where $\rho _m^T\left( \Phi \right) $ is the transposition of $\rho _m\left(
\Phi \right) .$ To prove this we need to verify three things. First, each $%
M_m$ is a positive operator and $\sum_{m=0}^{n-1}M_m=I_d.$ This is obvious
from Eq. (3). Second, when Alice implements this POVM measurement the
probability of an outcome $m$ is $p_m\left( \Phi \right) .$ This probability
is calculated as follows. $\left\langle \Psi _{AB}\right| M_m\left| \Psi
_{AB}\right\rangle =p_m\left( \Phi \right) tr\rho _m^T\left( \Phi \right)
=p_m\left( \Phi \right) .$ Third, when the outcome is $m$ the resultant
state of system $B$ is $\rho _m\left( \Phi \right) .$ This state is
calculated as follows. $\frac 1{p_m\left( \Phi \right) }tr_A\left( M_m\left|
\Psi _{AB}\right\rangle \left\langle \Psi _{AB}\right| \right) =\rho
_m\left( \Phi \right) .$

Given the unitary operations $\{u_m\}_{m=0}^{n-1}$, we can find an ensemble
of states that satisfy Eq. (3). If the number of states in the ensemble is
less than $n$ the RSP protocol is useless. We are interested in what
ensemble of states can be remotely prepared by using a given shared
entanglement resource. When we say an ensemble of states can be remotely
prepared, we mean that we can find a set of operators $\left\{ u_m\right\}
_{m=0}^{n-1}$ that satisfy Eq. (3) for any state in this ensemble.

In Eq. (3), it is obvious that $n\geqslant d$ must be satisfied. We have
investigated RSP protocols with $n=d$. These are faithful RSP protocols
using minimum classical bits.

{\it Theorem 1. }Suppose Alice and Bob have shared an entangled state in Eq.
(1). The ensemble of states 
\begin{equation}
\left\{ \left| \Phi \right\rangle =\sum_{j=0}^{d-1}\alpha _je^{i\varphi
_j}\left| j\right\rangle ,\forall \varphi _j\right\}   \eqnum{4}
\end{equation}
can be remotely prepared by using $\log d$ bits from Alice to Bob and their
previously shared entangled state, where $\left\{ \alpha _j\right\}
_{j=0}^{d-1}$ and $\left\{ \varphi _j\right\} _{j=0}^{d-1}$ are known to
Alice. Particularly, if Alice and Bob share a maximally entangled state, we
get the same results as those in Refs. \cite
{Pati/PRA63/014302,Zeng/PRA65/022316}.

{\it Proof. }We present an explicit method in which Alice prepared the
ensemble of states.

Suppose the state Alice wants Bob to prepare is 
\begin{equation}
\left| \Phi \right\rangle =\sum_{j=0}^{d-1}\alpha _je^{i\varphi _j}\left|
j\right\rangle ,  \eqnum{5}
\end{equation}
where $\left\{ \varphi _j\right\} _{j=0}^{d-1}$ is known to Alice. First,
Alice transforms locally the shared entangled state into 
\begin{equation}
\left| \Psi _{AB}\right\rangle =\sum_{i=0}^{d-1}\alpha _ie^{i\varphi
_i}\left| i\right\rangle \left| i\right\rangle ,\alpha
_i>0,\sum_{i=0}^{d-1}\alpha _i^2=1,  \eqnum{6}
\end{equation}
and then she performs a projective measurement on her system $A$ with the
measurement operators 
\begin{equation}
\left\{ P_m=\frac 1d\left( \sum_{j=0}^{d-1}e^{i\frac{2\pi }dmj}\left|
j\right\rangle \right) \left( \sum_{j=0}^{d-1}e^{-i\frac{2\pi }d%
mj}\left\langle j\right| \right) \right\} _{m=0}^{d-1}.  \eqnum{7}
\end{equation}
The measurement result, (supposed to be $m$ ), will be sent to Bob. When Bob
receives the message $m,$ he performs the corresponding unitary operation 
\begin{equation}
u_m=\sum_{j=0}^{d-1}e^{i\frac{2\pi }dmj}\left| j\right\rangle \left\langle
j\right|  \eqnum{8}
\end{equation}
on his system $B$ to transform the system $B$ into state (5). Q.E.D.

Obviously we have the following corollary.

{\it Corollary. }Suppose Alice and Bob have shared an entangled state in Eq.
(1). The ensemble of states $\left\{ v\left| \Phi \right\rangle =v\left(
\sum_{j=0}^{d-1}\alpha _je^{i\varphi _j}\left| j\right\rangle \right)
,\forall \varphi _j\right\} $ can be remotely prepared by using $\log _2d$
bits from Alice to Bob and their previously shared entangled state, where $v$
is an arbitrary unitary operators in $d$-dimensional Hilbert space

In the two-dimensional case the corollary shows that qubits chosen from the
same circle with radius $\sqrt{1-\left( \alpha _0^2-\alpha _1^2\right) ^2}$
on a Bloch sphere can be remotely prepared by using one classical bit from
Alice to Bob and their previously shared entangled state. Theorem 2 proves
that we have found all the ensembles in the two-dimensional case.

{\it Theorem 2. }Suppose Alice and Bob have shared an entangled state 
\begin{equation}
\left| \Psi _{AB}\right\rangle =\alpha _0\left| 0\right\rangle \left|
0\right\rangle +\alpha _1\left| 1\right\rangle \left| 1\right\rangle ,\alpha
_0,\alpha _1>0,\alpha _0^2+\alpha _1^2=1.  \eqnum{9}
\end{equation}
If there is an ensemble of states that can be remotely prepared by using one
bit from Alice to Bob and their previously shared entangled state, this
ensemble must be in the form 
\begin{equation}
\left\{ v\left| \Phi \right\rangle =v\left( \alpha _0\left| 0\right\rangle
+\alpha _1e^{i\varphi }\left| 1\right\rangle \right) ,\forall \varphi
\right\} ,  \eqnum{10}
\end{equation}
where $v$ is a unitary operator in two-dimensional Hilbert space.

{\it Proof. }If a state $\left| \Phi \right\rangle $ can be remotely
prepared, there should be unitary operators $u_0$ and $u_1,$ and
probabilities $p_0\left( \Phi \right) $ and $p_1\left( \Phi \right) $ which
satisfy the necessary and sufficient condition of RSP.

From Eq. (3) we have 
\begin{equation}
p_0\left( \Phi \right) u_0^{\dagger }\left| \Phi \right\rangle \left\langle
\Phi \right| u_0+p_1\left( \Phi \right) u_1^{\dagger }\left| \Phi
\right\rangle \left\langle \Phi \right| u_1=\alpha _0^2\left| 0\right\rangle
\left\langle 0\right| +\alpha _1^2\left| 1\right\rangle \left\langle
1\right| .  \eqnum{11}
\end{equation}
From Eq. (11) we can find that 
\[
\sqrt{p_0\left( \Phi \right) }\left( \frac 1{\alpha _0}\left| 0\right\rangle
\left\langle 0\right| +\frac 1{\alpha _1}\left| 1\right\rangle \left\langle
1\right| \right) u_0^{\dagger }\left| \Phi \right\rangle 
\]
and 
\[
\sqrt{p_1\left( \Phi \right) }\left( \frac 1{\alpha _0}\left| 0\right\rangle
\left\langle 0\right| +\frac 1{\alpha _1}\left| 1\right\rangle \left\langle
1\right| \right) u_1^{\dagger }\left| \Phi \right\rangle 
\]
form an orthonormal basis. So we can get 
\begin{equation}
\left\langle \Phi \right| u_0\left( \frac 1{\alpha _0^2}\left|
0\right\rangle \left\langle 0\right| +\frac 1{\alpha _1^2}\left|
1\right\rangle \left\langle 1\right| \right) u_1^{\dagger }\left| \Phi
\right\rangle =0.  \eqnum{12}
\end{equation}
It is the same as 
\begin{equation}
tr\left[ \left( \frac 1{\alpha _0^2}\left| 0\right\rangle \left\langle
0\right| +\frac 1{\alpha _1^2}\left| 1\right\rangle \left\langle 1\right|
\right) u_1^{\dagger }u_0\left( u_0^{\dagger }\left| \Phi \right\rangle
\left\langle \Phi \right| u_0\right) \right] =0.  \eqnum{13}
\end{equation}
We assume that 
\begin{equation}
u_1^{\dagger }u_0=\cos \frac{\theta _0}2I_2-i\sin \frac{\theta _0}2\left(
x_0\sigma _x+y_0\sigma _y+z_0\sigma _z\right)   \eqnum{14}
\end{equation}
and 
\begin{equation}
u_0^{\dagger }\left| \Phi \right\rangle \left\langle \Phi \right| u_0=\frac 1%
2\left( I_2+x\sigma _x+y\sigma _y+z\sigma _z\right) ,  \eqnum{15}
\end{equation}
where $\sigma _x,\sigma _y,$ and $\sigma _z$ are Pauli operators.

Substituting Eqs. (14) and (15) into Eq. (13), we get 
\begin{equation}
\cos \frac{\theta _0}2+z\left( \alpha _1^2-\alpha _0^2\right) \cos \frac{%
\theta _0}2+\left( \alpha _1^2-\alpha _0^2\right) \left( x_0y-y_0x\right)
\sin \frac{\theta _0}2=0,  \eqnum{16}
\end{equation}
\begin{equation}
\sin \frac{\theta _0}2\left[ \left( \alpha _1^2-\alpha _0^2\right)
z_0+x_0x+y_0y+z_0z\right] =0.  \eqnum{17}
\end{equation}
Because $\left| \Phi \right\rangle $ is a pure state, so 
\begin{equation}
x^2+y^2+z^2=1.  \eqnum{18}
\end{equation}

The common solutions of Eqs. (16)-(18) represent the ensemble of states that
can be remotely prepared. Generally, Eqs. (16) and (17) represent two planes
and Eq. (18) represents a sphere. If Eqs. (16) and (17) represent two
different planes there are at most two common solutions of Eqs. (16)-(18),
which are trivial. So we should seek the appropriate $u_1^{\dagger }u_0,$
which ensures that Eqs.(16) and (17) represent the same plane. The
requirement that Eqs. (16) and (17) represent the same plane leads to the
following results 
\begin{equation}
\theta _0=\pi ,x_0=y_0=0,z=\alpha _0^2-\alpha _1^2,\text{ when }\alpha
_0\neq \alpha _1,  \eqnum{19}
\end{equation}
\begin{equation}
\theta _0=\pi ,x_0x+y_0y+z_0z=0,\text{ when }\alpha _0=\alpha _1.  \eqnum{20}
\end{equation}

Eqs. (15), (19), and (20) show that qubits chosen from the same circle with
radius $\sqrt{1-\left( \alpha _0^2-\alpha _1^2\right) ^2}$ on a Bloch sphere
can be remotely prepared by using one classical bit from Alice to Bob and
their previously shared entangled state. This result is the same as the
corollary in two-dimensional case. Q.E.D.

\section{Remote Preparation of a general pure quantum state}

Now we turn to investigate RSP protocols which do not use minimum classical
bits. We will show that any pure quantum state can be faithfully remotely
prepared by using finite classical bits and a nonmaximally entangled state.
To prove this result, we first prove the following lemma.

{\it Lemma.} Suppose Alice and Bob have shared an entangled state in Eq.
(1). The ensemble of states 
\begin{equation}
S=\left\{ \left| \Phi \right\rangle =\sum\limits_{j=0}^{d-1}\beta
_je^{i\varphi _j}\left| j\right\rangle ,\forall \varphi _j,\left( \alpha
_j^2\right) _{j=0}^{d-1}\prec \left( \beta _j^2\right) _{j=0}^{d-1}\right\} 
\eqnum{21}
\end{equation}
known to Alice can be remotely prepared by using $\log _2d+m$ bits from
Alice to Bob and their previously shared entangled state, where $m$ is equal
to $\log _2d$ when they initially shared a maximal entangled state,
otherwise $m$ is equal to $\log d!$. The symbol $\prec $ has the same
meaning as that in \cite{Nielsen/PRL83/436}.

{\it Proof : }We can accomplish our remote state preparation by two steps.

{\it Step 1}. Alice and Bob transform their shared entangled state into 
\begin{equation}
\left| \Psi _{AB}\right\rangle =\sum_{i=0}^{d-1}\beta _i\left|
i\right\rangle \left| i\right\rangle ,\beta _i\geqslant
0,\sum_{i=0}^{d-1}\beta _i^2=1  \eqnum{22}
\end{equation}
by using $m$ bits from Alice to Bob \cite{Nielsen/PRL83/436}. For the final
entangled state has the same Schmidt basis as the original one, Bob can only
perform permutative operation to accomplish the transformation, which
indicates Bob need not know the final state \cite{Jensen/PRA63/062303}. The
total number of such operation is $d!,$ i.e. $m=\log _2d!$ will be enough.
Especially when the initially shared entangled state in Eq. (1) is a maximal
one, $m=\log _2d$ will be enough \cite{Lo /PRA62/012313}.

{\it Step 2}. According to Theorem 1, Alice and Bob use their new shared
entangled state in Eq. (22) to prepare the state 
\begin{equation}
\left| \Phi \right\rangle =\sum\limits_{j=0}^{d-1}\beta _je^{i\varphi
_j}\left| j\right\rangle   \eqnum{23}
\end{equation}
by using $\log _2d$ bits from Alice to Bob. Note that Alice does not receive
classical message from Bob, Alice can send $\log _2d+m$ bits together to Bob
and Bob performs the corresponding unitary operation to accomplish the
remote state preparation. Q.E.D.

In the above we have presented an ensemble of state $S$ that can be remotely
prepared by using finite classical bits communication. If we can find finite
unitary operations $\left\{ u_i\right\} _{i=1}^m$ such that $\left| \Phi
\right\rangle \in \bigcup\limits_{i=1}^m\left( u_iS\right) $ for any pure
state $\left| \Phi \right\rangle $, then we can claim that any pure state
can be remotely prepared by using finite classical bits communication from
Alice to Bob. Fortunately there exist such finite unitary operations
satisfying the condition. This result relies on Heine-Borel theorem \cite
{borel}.

{\it Theorem 3.} Suppose Alice and Bob have shared an entangled state in Eq.
(1). Any pure state of dimension $d$ can be remotely prepared by using
finite classical bits from Alice to Bob and their previously shared
entangled state.

{\it Proof. }From Heine-Borel theorem \cite{borel} we can conclude that the
set $A=\left\{ x\in C^d|\left\| x\right\| =1\right\} $ is compact \cite{norm}%
. Because the set $F=\left\{ B\left( x_0,r\right) |x_0\in A\right\} $ is an
open cover of $A$, where $B\left( x_0,r\right) =\left\{ x\in C^d|\left\|
x-x_0\right\| <r\right\} $, so $F$ admits a finite subcover. This means, for
any $r>0,$ there exists a finite $n_r\in N$ such that the set $G=\left\{
B\left( x_i,r\right) \cap A|x_i\in A,i=1,...,n_r\right\} $ is a cover of $A.$
That is to say $\bigcup\limits_{i=1}^{n_r}\left\{ B\left( x_i,r\right) \cap
A\right\} =A.$ Since unitary operations preserve the norm, any two elements
of $G$ can be connected by a unitary operation. There is a bijective map
between the set $A$ and the set of all pure state of dimension $d$ which
maps the state $\sum\nolimits_{i=0}^{d-1}\alpha _ie^{i\varphi _i}\left|
i\right\rangle $ to the point $\left( \alpha _0e^{i\varphi _0},...,\alpha
_{d-1}e^{i\varphi _{d-1}}\right) .$ So we can regard the state set $S$
presented in the lemma as a subset of $A$. We assume that $u_0$ is the image
of the state $\left| 0\right\rangle $. It can be easily verified that when $%
0<r<\min \left( \alpha _0,...,\alpha _{d-1}\right) ,$ the set $B\left(
u_0,r\right) \cap A$ is a subset of $S$. This means that there is a cover $G$
of $A$ which has finite element and each element can be generated by a
unitary operation performed on the subset $B\left( u_0,r\right) \cap A$ of $S
$. Note that $A$ represent all pure states and $S$ represent the state set
we given in the lemma, we can finish our proof. Q.E.D.

\section{Discussion And Summary}

In Sec. II, we have discussed the RSP protocols by using minimum classical
bits and we have found many ensembles of states that can be remotely
prepared by using minimum classical bits and the previously shared entangled
state. Any two such ensembles are connected by a unitary operation. In some
special cases, we can find more ensembles of states that can be remotely
prepared by using the same resource and the connection between them is not
necessarily a unitary operation. For example, when Alice and Bob share the
entangled state 
\begin{equation}
\left| \Psi _{AB}\right\rangle =\alpha \left| 00\right\rangle +\beta \left|
11\right\rangle +\alpha \left| 22\right\rangle +\beta \left| 33\right\rangle
,  \eqnum{24}
\end{equation}
the ensemble of states 
\begin{equation}
\left\{ \left| \Phi \right\rangle =\alpha \left| 0\right\rangle +e^{i\varphi
_1}\beta \left| 1\right\rangle +\alpha e^{i\varphi _2}\left| 2\right\rangle
+\beta e^{i\varphi _3}\left| 3\right\rangle ,\forall \varphi
_j,j=1,2,3\right\}   \eqnum{25}
\end{equation}
and the ensemble of states 
\begin{equation}
\left\{ \left| \Phi \right\rangle =\alpha \left| 0\right\rangle +e^{i\varphi
}\beta \left| 1\right\rangle ,\forall \varphi \right\}   \eqnum{26}
\end{equation}
can both be remotely prepared by using $2$ classical bits from Alice to Bob
and their previously shared entangled state. But these two ensemble of
states can not be connected by a unitary operation. Actually we can get the
ensemble of states in Eq. (26) by performing a quantum measurement \cite
{Chuang} with measurement operators 
\begin{equation}
E_0=\left| 0\right\rangle \left\langle 0\right| +\left| 1\right\rangle
\left\langle 1\right| ,E_1=\left| 0\right\rangle \left\langle 2\right|
+\left| 1\right\rangle \left\langle 3\right|   \eqnum{27}
\end{equation}
on the ensemble of states 
\begin{equation}
\left\{ \left| \Phi \right\rangle =\alpha \left| 0\right\rangle +e^{i\varphi
}\beta \left| 1\right\rangle +e^{i\psi }\left( \alpha \left| 2\right\rangle
+\beta e^{i\varphi }\left| 3\right\rangle \right) ,\forall \varphi ,\forall
\psi ,\right\}   \eqnum{28}
\end{equation}

In Sec. III, we have proved that any pure quantum state can be remotely
prepared by using finite classical bits and the previously shared
non-maximally entangled qubit states.

\begin{center}
{\bf Acknowledgments}
\end{center}

We thank Z. W. Zhou, Y. C. Wu, Y. J. Han and Y. Hu for their helpful advice.
We also thank the referees for their comments and helpful advice. This work
was funded by National Fundamental Research Program (2001CB309300), National
Natural Science Foundation of China.

\end{document}